\documentclass[preprint,amsmath,amssymb]{revtex4}
\usepackage{graphicx}
\usepackage{dcolumn}
\usepackage{bm}
\begin{document}
\title{Electromagnetic Casimir effect and the spacetime index of refraction}
\author{ B.~Nazari $^{a}$ \footnote{Electronic
address:~bornadel@khayam.ut.ac.ir}
and  M.~Nouri-Zonoz $^a$ \footnote{Electronic address:~nouri@theory.ipm.ac.ir, corresponding author} }
\address{$^{a}$ Department of Physics, University of Tehran, North Karegar Ave., Tehran 14395-547,
Iran.}

\begin{abstract}
In \cite{NouriNazari} we investigated the response of vacuum energy
to a  gravitational field by considering a Casimir apparatus in a
weak gravitational field. Our approach was based on a conjecture
involving the interpretation of spacetime as a refractive medium and
its effect on vacuum energy composed of virtual massless scalar
particles. There it was shown how the case of virtual photons as the
constituents of vacuum could be inferred from that of the massless
scalar field. Here we explicitly show how the same conjecture
applies to the electromagnetic vacuum composed of virtual photons.
Specifically we show that the boundary conditions imposed on the
components of the vector field, decomposed into two scalar fields, 
result in the same frequency shift for photons. Using the same decomposition
and employing our conjecture, we also calculate the electromagnetic energy 
density for the Casimir apparatus in a weak gravitational field.
\end{abstract}
\maketitle
\section{Introduction}
One of the places where the general theory of relativity and
quantum field theory meet is the notion of quantum
vacuum and its role in the formulation of the so called {\it cosmological constant problem}.
Indeed one could ask three fundamental questions concerning vacuum (zero-point) energy with 
descending level of importance. These are as follows:\\
1-Is zero-point energy real? this is an old question which has been raised recently in the wake of the
cosmological constant problem \cite{Jaffe}.\\
2-If vacuum energy is taken to be real, does it gravitate? Indeed an implicit affirmative answer to this question has
naturally led to the cosmological constant problem \cite{weinberg}, while a negative answer
(through a consistent mechanism)
has been sought as part of a possible resolution to the same problem \cite {paddy}.\\
3-If vacuum energy is real and gravitates, how does it respond to a gravitational field?\\
In our previous study we concentratred on this last question by considering a Casimir
apparatus in a weak gravitational field and studying the effect of the gravitational field on
Casimir energy as a macroscopic manifestation of vacuum energy.
The same problem has been addressed in \cite{Fulling} by considering the variation in the
gravitational energy in a static, weak gravitational field represented by the Fermi metric \footnote{Note that the convention used here for the indices is different from that in our previous paper \cite{NouriNazari} such that now Roman indices run from $1$ to $3$ while Greek ones run from $0$ to $3$.}
\begin{eqnarray}\label{metric1}
g_{00} = (1 +  2gz)\;\;\;\; , \;\;\;\; g_{ij} =
-\delta_{ij}\;\;\;\;, \;\;\;\; i,j = 1,2,3,
\end{eqnarray}
leading to the following force acting on
the {\it renormalized casimir energy},
\begin{eqnarray}\label{force0}
{F} = -g {\cal E}_{Casimir} = g\frac{\pi^2 {\hslash}}{720 \ell^3 c},
\end{eqnarray}
in which $\ell$ is the plate separation. The above equation states that the equivalent gravitational mass of the Casimir energy is
nothing but the energy itself. As seen, the
{\it upward} force is a very small effect compared to the gravitational force acting
on the plates. In \cite{NouriNazari} we investigated
the same setup  using a conjecture involving the
interpretation of spacetime as a refractive medium and its effect on vacuum energy constituted of
virtual massless particles. To make this article self-sufficient and
to explain the conjecture in a more exact wording it is stated below:\\
{\bf conjecture}:
{\it Accepting the reality of electromagnetic zero-point energy as the propagation of virtual photons ,
their propagation, as in the case of real photons, should be influenced by the presence of a gravitational
field through the interpretation of modified Fermat's principle in a curved spacetime. This influence,
in the presence of boundaries, is encoded in the frequency shift of virtual massless particles induced
by the spacetime index of refraction.}\\
Attributing an index of refraction to a spacetime means that if the
allowed photon wavelengths are somehow fixed, e.g. through the
presence of boundaries such as in the case of the Casimir apparatus,
then the effect of refractive index would be the change in the
photon frequencies from $\omega_k$ to $\frac{\omega_k}{n}$. In
\cite{NouriNazari} we showed the consistency of the results obtained
through the above conjecture  by explicitly solving the Klein-Gordon
equation for a massless scalar field in a Casimir apparatus in Fermi
metric (\ref{metric1}) representing the weak background spacetime.
To extend our results to the case of electromagnetic Casimir energy
(i.e to the case of  virtual photons as the constituents of vacuum)
we employed a result from quantum field theory in flat spacetime according to
which the energy content of (virtual) photons could be inferred from that of
massless scalar fields, on account of the fact that the
electromagnetic field has twice as many modes. Here we show the
consistency of our conjecture, applied explicitly to the
electromagnetic vacuum energy, by the well know method of decomposing the  
vector potential (field) into (gradiants of) two scalar fields. The outline of 
the paper is as follows. In the next section we give a brief review of the scalar field results
of \cite{NouriNazari} and in section III the solution to the Maxwell
equations in curved background and the decomposition of the vector
potential into two scalar fields will be discussed. In section IV,
application of the boundary conditions on the vector potential mode
functions will be shown to result in the same frequency shift as in
the case of massless scalar fields. In section V we calculate the energy
density of the electromagnetic field in Fermi metric by employing our conjecture 
on the constituent scalar fields into which the vector potential is decomposed.
The results are summarized and discussed in the last section.
\section{Scalar Casimir effect in a weak gravitational field}
The massless scalar field satisfying the Klein-Gordon equation in a curved background
\begin{eqnarray}\label{FE}
\partial_\mu[\sqrt{-\mathfrak{g}} g^{\mu\nu} \partial_\nu \Phi(x^c)]= 0
\;\;\;\;\;\;,\;\;\; \mathfrak{g}\equiv {\rm det}g^{\mu\nu}.
\end{eqnarray}
is quantized by imposing the following commutation relations
\begin{eqnarray}\label{eq002}
(\Phi_i(x) , \Phi_j(x)) = \delta_{ij}\delta({\bf k}_i - {\bf k}_j).
\end{eqnarray}
where the scalar product is defined as \cite{Birrell}
\begin{eqnarray}\label{eq001}
(\Phi_1 , \Phi_2) =
-i\int_{\Sigma}\Phi_1(x){\overleftrightarrow{\partial}}_a
\Phi_2^*(x)[-\mathfrak{g}_{\Sigma}(x)]^{\frac{1}{2}}n^a d\Sigma,
\end{eqnarray}
For Casimir apparatus in Fermi metric (\ref{metric1}) the solution
to equation (\ref{FE}) was given as follows \cite{NouriNazari}:
\begin{eqnarray}\label{SF0}
\Phi(x) = C e^{-i\omega t} e^{ik_x x}e^{ik_y y}Z(z),
\end{eqnarray}
in which $C$ is the normalization constant
\begin{eqnarray}\label{C}
C^2 = \frac{1}{2(2\pi)^2 \omega} [\int_0^\ell
\frac{Z^2(z)}{\sqrt{g_{00}}} dz]^{-1}.
\end{eqnarray}
and (to the first order in $gz$) ,
\begin{eqnarray}\label{eq16}
Z(z) \approx Z_0 (1+\alpha_0 z) \sin (A(z)+\theta_0)
\end{eqnarray} where  $\alpha_0=-[\frac{g}{2} + \frac{a}{4b}]$ , $b=\omega^2 -
k_{\perp}^2$ , $a = -2g\omega^2$ and  $A(z)$ is given by
\begin{eqnarray}\label{A(z)-function}
A(z)=\int_0^z \sqrt{S(z')}dz'\;\;\;,\;\;\;S =
\frac{\omega^2}{-\mathfrak{g}} - k^2_\perp +
\frac{3}{4}(\frac{g}{\mathfrak{g}})^2.
\end{eqnarray}
with $k_{\perp}^2 = k^2_x + k^2_y$. Applying the Dirichlet boundary
condition on plates we have, up to first order in $g$
\begin{eqnarray}\label{omegazero}
\omega\approx (1+\frac{g\ell}{2} )\omega_0.
\end{eqnarray}
This result, as we have proposed in \cite{NouriNazari}, can be obtained through the following
frequency shift
\begin{eqnarray}\label{eq9a}
\omega = \frac{\omega_0}{n}
\end{eqnarray}
in which $n$ is the spacetime index of refraction, which for static spacetimes including Fermi
metric is given by $\frac{1}{\sqrt{g_{00}}}$. Therefore
\begin{eqnarray}\label{eq9}
\omega = \omega_0 (g_{00})^{\frac{1}{2}}=
(1+2g<z>)^{\frac{1}{2}}\omega_0 \approx (1+g<z>)\omega_0
\end{eqnarray}
where $<z> = \frac{\ell}{2}$ leads to the same result as in (\ref{omegazero}). This is a direct consequence of
the application of our conjecture to virtual massless  scalar particles as the elements of vacuum in a Casimir apparatus placed in a weak gravitational field.
\section{Electromagnetic field in the Fermi metric}
Source free Maxwell field equations in the Lorentz gauge and in a curved
background read as follows \cite{Birrell}
\begin{eqnarray}\label{ME}
\square A^{\mu} + R^\mu_\nu A^\nu=0 \;\;\;\   , \;\;\;
\nabla_\mu A^{\mu}=0 \;\;\;\   , \;\;\;\mu = 0,1,2,3.
\end{eqnarray}
where $\square$ is the corresponding Dalambertian operator in curved
spacetime. It was noted in \cite{NouriNazari} that  both the Riemmann and  Ricci tensors  are
of order $g^2$ in Fermi metric and hence does not appear in the above equation within our approximation which is
up to the first order in $g$. Hence, in the absence of curvature $R^\mu_\nu$, every component
of the above equation could be thought of as a Klein-Gordon
equation for that component of the vector field but now with the extra Lorentz condition as a differential
relation among the components.
The Lorentz condition and the hidden gauge freedom in the above equations reduce the number of photon degrees of freedom to the
two physical ones corresponding to transverse photons. Based on this fact
there are two approaches to the quantization of photons. In the first approach all the degrees of
freedom (including the scalar and longitudinal ones) are kept in the quantization procedure and then the Lorentz
condition is employed in a weaker form leading to a subsidiary condition on the linear combination of
longitudinal and scalar photons. As a consequence of this subsidiary condition, only transverse photons contribute in
the expectation value of physical observables such as energy. This approach  known as the Gupta-Bleuler formalism respects the
covariance of the theory \cite{Mandl}. In the second approach one first reduces the degrees of
freedom to the physical ones and then quantizes the theory. Here we employ the second approach \cite{Deutch} in the context of
photon quantization in a weak static gravitational field represented by the Fermi metric (\ref{metric1}).
Starting with the covariant gauge transformation
\begin{eqnarray}\label{G-Fixing}
A'_{\mu}=A_{\mu} + \nabla_\mu \Lambda
\end{eqnarray}
along with the fact that the vector potential  should be of the form (by the symmetries of the Fermi metric)
\begin{eqnarray}\label{A-form}
A_\mu^{(\lambda)}=
f_\mu^{(\lambda)}(z) e^{-i\omega t} e^{ik_x x}e^{ik_y y} \;\;\;\ , \;\;\;\; (\lambda = 0,...,3)
\end{eqnarray}
would allow us to break the Lorentz condition to the following two separate conditions \footnote{We use the convention
employed in \cite{Deutch} where $a,b,c,d = 0,3$ and $i,j,k,l = 1,2$ and this should not be confused with 
our convention for the use of Roman indices elsewhere in the paper. }
\begin{eqnarray}\label{A-form1}
\nabla^a A_a=0 \;,\;a=0,3\equiv(t,z) \;\   , \;\;\;
\nabla^i A_i=0 \;,\;i=1,2\equiv(x,y).
\end{eqnarray}
This could be achieved by choosing $\Lambda = - \frac{ik^iA_i}{k^2}$ through the gauge freedom \cite{Deutch}.
The above separation implies the existence of scalar functions $\phi$ and $\psi$ such that
\begin{eqnarray}\label{G-Breaking}
A_i=\varepsilon_{ij}\nabla^j\phi \;\   , \;\;\;
A_a=\varepsilon_{ab}\nabla^b\psi.
\end{eqnarray}
in which
\begin{eqnarray}\label{spinor-matrixes}
\epsilon_{ij}=\left(
                \begin{array}{cc}
                   0 & 1 \\
                   -1 & 0 \\
                \end{array}
              \right) \;\; ,\;\;  \epsilon_{ab}=\left(
                                                   \begin{array}{cc}
                                                       0 & 1+gz \\
                                                       -1-gz & 0 \\
                                                   \end{array}
                                                \right).
\end{eqnarray}
Obviously on account of relations (\ref{A-form}), (\ref{G-Breaking})
and the wave equation (\ref{ME}) both functions $\psi$ and $\phi$
satisfy the Klein-Gordon equation (see appendix A). Using equations
(\ref{G-Breaking}) and (\ref{spinor-matrixes}) recasts the situation to that of the
scalar field problem, leading to the following form for the
quantized vector potential \cite{Bimonte}
\begin{eqnarray}\label{A-expansion}
{A}_{\mu}= (A_a , 0) + (0 , A_i) =\sum_{r=1}^\infty \int \frac{d^2 {\bf k}}{k(2
\pi)^2}\sum_{\lambda=1}^4[{ A}_{r{\bf
k}\mu}^{(\lambda)}(x)a_{r\lambda}({\bf k})+{A}_{r{\bf k}
\mu}^{(\lambda)*}(x)a_{r\lambda}^*({\bf k})],
\end{eqnarray}
where
\begin{eqnarray}\label{A-mode1}
{A}_{r{\bf k} \mu}^{(0)}(x) =(\nabla_a,{\bf 0})\, \phi_{r{\bf
k}}(x)=(-i\omega\phi_{r{\bf k}}(x),\partial_z\phi_{r{\bf
k}}(x),0,0),
\end{eqnarray}
\begin{eqnarray}\label{A-mode2}
{ A}_{r{\bf k}\mu}^{(1)}(x) =(p_a,{\bf 0})\, \psi_{r{\bf
k}}(x)=((1+gz)\partial_z\psi_{r{\bf k}}(x),-i(1-gz)\omega\psi_{r{\bf
k}}(x),0,0),
\end{eqnarray}
\begin{eqnarray}\label{A-mode3}
{A}_{r{\bf k} \mu}^{(2)}(x) =({\bf 0},p_i)\, \phi_{r{\bf k}}(x)=(0,0,ik_y
\phi_{r{\bf k}}(x),-ik_x \phi_{r{\bf k}}(x))
\end{eqnarray}
\begin{eqnarray}\label{A-mode4}
{A}_{r{\bf k} \mu}^{(3)}(x) =({\bf 0},\nabla_i)\, \phi_{r{\bf
k}}(x)=(0,0,ik_x \phi_{r{\bf k}}(x),ik_y \phi_{r{\bf k}}(x)).
\end{eqnarray}
are the linearly independent modes of the vector potential (see appendix A) with the operators  
$p_a=\epsilon_{ab} \nabla^b$ and  $p_i=\epsilon_{ij}
\nabla^j$ already introduced in (\ref{G-Breaking}). 
It should also be noted that due to the
presence of the boundary conditions along the $z-$axis, the integration over $k_z$ is now replaced 
with the summation over the extra index $r$, representing the $r$-th
harmonic along the $z$-axis. Confining to the physical degrees of freedom ($\lambda = 1,2$) 
and substituting equations (\ref{A-mode2}) and (\ref{A-mode3}) (along with their conjugates) into (\ref{A-expansion}) we end up with
\begin{eqnarray}\label{A-expansionA}
{A}_{\mu}= (p_a , 0)\sum_{r=1}^\infty \int \frac{d^2 {\bf k}}{k(2
\pi)^2}[{\psi}_{r{\bf k}}(x)a_{r1}({\bf k})+{\psi}_{r{\bf k}}^{*}(x)a_{r1}^*({\bf k})] \cr\cr
+ \; (0 , p_i)\sum_{r=1}^\infty \int \frac{d^2 {\bf k}}{k(2
\pi)^2} [\phi_{r{\bf k}}(x)a_{r2}({\bf k})+{\phi}_{r{\bf k}}^{*}(x)a_{r2}^*({\bf k})] 
\end{eqnarray} 
which is the same as equation (28) in reference \cite{Deutch} for $a_{r1}({\bf k})\equiv a_r({\bf k})$ 
and $a_{r2}({\bf k})\equiv b_r({\bf k})$.
In the next section it is shown how
one can employ the mode sums used in \cite{NouriNazari} to obtain the frequency shift on the scalar 
fields into which the vector field (potential) is decomposed as above. That will consequently lead to
the frequency shift on the vector potential itself.
\section{Boundary conditions on the vector potential and the photon frequency shift}
Now that it is shown how a vector field could be {\it formally} decomposed into two scalar fields
satisfying the Klein-Gordon equation, the next task is to find out how the well known boundary 
conditions on the vector potential 
are reduced to those on the scalar fields $\phi$ and $\psi$ into which it is decomposed. To find these boundary conditions 
we start with the boundary conditions on the electric and magnetic fields on plates
(${\partial M}$) which could be covariantly encoded into the following formula
\begin{eqnarray}\label{Ftensor-boundary}
n^\mu {\tilde F}_{\mu\nu}\mid_{\partial M}=0
\end{eqnarray}
with $n^\mu = (0, 1, 0, 0)$ \footnote{Note that here $\mu = 0,3,1,2$,
in line with our separation in  (\ref{A-form1}).} unit vector normal
to the boundary and ${\tilde F}_{\mu\nu}$ the dual Maxwell tensor. These covariant boundary conditions on the field
tensor are equivalent to the well known boundary conditions on the
electric and magnetic components of the electromagnetic tensor
\begin{eqnarray}
{\bf E}_\parallel\mid_{\partial M}={\bf B}_\perp\mid_{\partial M}=0
\end{eqnarray}
Using the definitions of Maxwell tensor in terms of the vector potential, the boundary conditions  (\ref{Ftensor-boundary})
on the  derivatives of the vector potential read as  
\begin{eqnarray}\label{A-boundary1}
A_{t,x}^{(\lambda)}\mid_{\partial
M}=A_{x,t}^{(\lambda)}\mid_{\partial
M}\;\;\;,\;\;A_{t,y}^{(\lambda)}\mid_{\partial
M}=A_{y,t}^{(\lambda)}\mid_{\partial
M}\;\;,\;\;A_{x,y}^{(\lambda)}\mid_{\partial
M}=A_{y,x}^{(\lambda)}\mid_{\partial M}
\end{eqnarray}
from which, by taking into account the general form of the vector potential 
(\ref{A-form}) and the fact that the boundaries are at $z= constant$, we end up with
\begin{eqnarray}\label{A-boundary2}
A_t^{(\lambda)}\mid_{\partial M}=A_x^{(\lambda)}\mid_{\partial
M}=A_y^{(\lambda)}\mid_{\partial M}=0
\end{eqnarray}
This could be easily obtained in the coulomb gauge where one can always choose 
$A_0^{(\lambda)} = 0$ everywhere including on the boundary. Upon substitution of this choice in the first two
equations of (\ref{A-boundary1}) and taking into account the general form (\ref{A-form})
we end up with
\begin{eqnarray}\label{A-boundary2aa}
-i\omega A_{x}^{(\lambda)}\mid_{\partial M}= -i\omega A_{y}^{(\lambda)}\mid_{\partial M} = 0
\end{eqnarray}
in other words the components of the vector potential tangential to the boundary vanish. It remains the boundary condition on the $z$-component of the
vector potential to be specified and this will be done by employing
the Lorentz gauge condition $A^\mu_{\; ;\mu} = 0$ on the boundary
\begin{eqnarray}\label{A-boundary2a}
\partial_\mu(\sqrt{-\mathfrak{g}}g^{\mu\nu}A_\nu^{(\lambda)})\mid_{\partial M}=0
\end{eqnarray}
Since the boundary is at $z=constant$ the above relation by virtue of the relations 
(\ref{A-boundary2}) reduces to
\begin{eqnarray}\label{A-boundary2b}
\partial_z(\sqrt{-\mathfrak{g}}g^{zz}A_z^{(\lambda)})\mid_{\partial M}=0
\end{eqnarray}
which on using the fact that, to the first order in $g$,  $\sqrt{-\mathfrak{g}}\approx (1+gz) $  becomes
\begin{eqnarray}\label{A-boundary3}
(\partial_z A^{(\lambda)}_z+gA_z^{(\lambda)})\mid_{\partial M}=0
\end{eqnarray}
Imposing the boundary conditions (\ref{A-boundary2}) and
(\ref{A-boundary3}) on the vector potential mode functions
(\ref{A-mode1})-(\ref{A-mode4}), one  arrives at the following
boundary conditions on the scalar potentials $\phi$  and $\psi$ (for
all values of $\lambda = 0,1,2,3 $)
\begin{eqnarray}\label{final-boundary1}
\phi |_{\partial M}=\partial_z\psi|_{\partial M}=0
\end{eqnarray}
\begin{eqnarray}\label{final-boundary2}
\partial_z^2 \phi\mid_{\partial M} = -g\partial_z \phi|_{\partial M}
\end{eqnarray}
For consistency check purposes, starting from the first equation of
(\ref{final-boundary1}) and the Klein-Gordon equation for $\phi$, it
is an easy task to show that the equation (\ref{final-boundary2})
will be obtained. In fact we recall that in \cite{NouriNazari} it
was shown that the Klein-Gordon equation (\ref{SF0}) reduces to the
following equation for function $Z(z)$
\begin{eqnarray}\label{KG-linear}
Z''(z)+P(z)Z'(z)+Q(z)\emph{Z(z)}=0
\end{eqnarray}
with
\begin{eqnarray}\label{KG-linear1}
P(z)=\partial_z \ln\sqrt{-\mathfrak{g}}\;\;,\
;\;Q(z)=-(k_\perp^2+\omega^2g^{00})
\end{eqnarray}
in which $'\equiv \partial_z$. Recalling that the function $Z(z)$ is that part of the scalar field on which 
the boundary condition should be satisfied, use of $\phi |_{\partial
M}=0$ in (\ref{KG-linear}) immediately returns (\ref{final-boundary2}). Similar analysis
from the second equation of (\ref{final-boundary1}) and the
Klein-Gordon equation for $\psi$ will result in
\begin{eqnarray}\label{final-boundary3}
\partial_z^2 \psi\mid_{\partial M} =(k_\bot^2+g^{00}\omega^2)\psi|_{\partial M}
\end{eqnarray}
So up to now, the boundary conditions on the vector potential are
recast into the equations (\ref{final-boundary1}) ,
(\ref{final-boundary2}) and (\ref{final-boundary3}). Since the
scalar field $\phi$ inherits the same frequency $\omega$ from its
parent vector field in (\ref{A-form}), from our previous study
\cite{NouriNazari} it is clear that the first condition in
(\ref{final-boundary1}), $\phi\mid_{\partial M}=0$, leads to the
frequency shift
\begin{eqnarray}\label{shift}
\omega=\omega_0(1+\frac {g\ell}{2}).
\end{eqnarray}
It is expected that the second condition in (\ref{final-boundary1})
, $\partial_z \psi\mid_{\partial M}=0$, should also lead to the same
frequency shift if our formulation is consistent. This is
demonstrated to be the case in what follows. To proceed looking for
the Neumann boundary condition on the scalar field $\psi$, from
(\ref{FE}), (\ref{SF0}), (\ref{eq16}) and (\ref{A(z)-function}) 
we have up to the first order in $g$
\begin{eqnarray}\label{Z-prime1}
&& Z'(0)=Z_0\{\alpha_0\sin(\theta_0)+\sqrt{b}\cos(\theta_0)\}=0 \\
&& Z'(\ell)=Z_0\{\alpha_0\sin(A(\ell)+\theta_0)+(1+\alpha_0
\ell)\sqrt{b}(1+\frac{a}{2b}\ell)\cos(A(\ell)+\theta_0)\}=0
\end{eqnarray}
leading to
\begin{eqnarray}\label{E-components3}
&& \tan(\theta_0)=-\frac{\sqrt{b}}{\alpha_0}\\
&&
\tan(A(\ell)+\theta_0)=-\frac{\sqrt{b}}{\alpha_0}(1+(\alpha_0+\frac
{a}{2b}\ell))
\end{eqnarray}
which in turn result in
\begin{eqnarray}\label{shift3}
\tan(A(\ell))=-\frac{-\alpha_0
\sqrt{b}(\alpha_0+\frac{a}{2b})\ell}{\alpha_0^2+b(1+(\alpha_0+\frac
{a}{2b}\ell))}
\end{eqnarray}
Now since both $\alpha_0$ and $a$ are proportional to $g$  the numerator of the fraction on the right hand side is of order
$\emph{O}(g^2)$ and so up to first order we have
\begin{eqnarray}\label{shift-final}
\tan(A(\ell))=0\;\;\Rightarrow\;\; \sin(A(\ell))=0\;\;\Rightarrow\;\;A(\ell)=\int_0^l \sqrt{S(z')}dz'=n\pi
\end{eqnarray}
This is exactly the equation $(36)$ of \cite{NouriNazari} which was
shown to lead to the frequency shift  in (\ref{shift}). In other
words both the Neumann boundary condition on $\psi$ and the Dirichlet
boundary condition on  $\phi$ led to the same frequency shift on
virtual photons as the constituents of quantum vacuum in a Casimir
apparatus in a weak gravitational field.
\section{Calculation of the energy density}
Since we have formally two scalar fields replacing an
electromagnetic field, it may be expected that the energy content to
be doubled compared to the case of Casimir vacuum composed of
virtual massless scalar particles. This expectation is not a trivial
one since the decomposition of the electromagnetic field into two
scalar fields is such that not only they are defined in the two
totally different sectors of the space time but also with different
boundary condition in Fermi metric. To further reinforce our
conjecture, in this section we explicitly calculate the
energy-momentum tensor of the electromagnetic field to find the
energy of the photon field without any resort to the Green functions
method frequently employed in the literature. To do so first we
discuss the energy momentum tensor in Fermi metric in the classical
context by noting that the Maxwell energy-momentum tensor is given
by
\begin{eqnarray}\label{ET1}
T^{\mu\nu} = \frac{1}{4} g^{\mu\nu} F^{\lambda\tau} F_{\lambda \tau} - F^{\mu\tau} F^{\nu}_{\tau}.
\end{eqnarray}
Using Fermi metric and the fact that $F_{ij} = \epsilon_{ijk}
B^k\;\;\;\; (i,j,k=1,2,3) $ it could be shown that
\begin{eqnarray}\label{ET2}
F^{\lambda\tau} F_{\lambda \tau} = - 2 g^{00} E^2 + \epsilon^{ij}_{k^\prime}\epsilon_{ijk} B^k B^{k^\prime} = 2 (B^2 - g^{00} E^2),
\end{eqnarray}
from which we obtain
\begin{eqnarray}\label{ET3}
T^{00}  = \frac{1}{2} {g^{00}}^2 (E^2 + g_{00} B^2),
\end{eqnarray}
where we have used the fact that $g^{00} = \frac{1}{g_{00}}$. Therefore the energy density is given by
\begin{eqnarray}\label{ET4}
T_{00}  = \frac{1}{2} (E^2 + g_{00} B^2).
\end{eqnarray}
Note that the electromagnetic fields appearing in the above relation are those satisfying the Maxwell equation in a weak gravitational field
represented by the Fermi metric. To calculate the vacuum energy density we use the above classical result as the only contribution to the time-time component of energy-momentum tensor but it should be kept in mind that
in the quantum field theoretic treatment of a vector field (spin 1) in a curved background there are other contributions to the energy-momentum tensor
corresponding to the gauge breaking and ghost terms in the Lagrangian \cite{Birrell}. In the case of photons it has already been shown, through point splitting regularization, that these terms will lead to vanishing contribution to the vacuum expectation value of energy-momentum tensor in a weak gravitational field represented by the Fermi metric \cite{Bimonte}.
Now to compute the expectation value $<0|T_A^{00}|0>$ we start with the calculation of the electromagnetic fields in terms of the components of the Maxwell tensor. The electromagnetic fields are
\begin{eqnarray}\label{E&B-vectorform}
B_{i}=\epsilon_{ijk}F^{jk} \;\;,\;\;
\mathbf{E}=(F_{01},F_{02},F_{03}),
\end{eqnarray}
which have the following squared values in terms of the components of the Maxwell tensor in the coordinates chosen in section III,
\begin{eqnarray}\label{E&B-square}
&&|\mathbf{E}|^2=|F_{\tau z}|^2+|F_{\tau
x}|^2+|F_{\tau y}|^2 \\
&&|\mathbf{B}|^2=|F_{z x}|^2+|F_{zy}|^2+|F_{x y}|^2.
\end{eqnarray}
Using the expansion (\ref{A-expansion}) of the vector potential in terms of the general mode functions (\ref{A-form}) we can calculate the above components in terms of their decomposition into scalar fields $\phi$ and $\psi$ introduced in (\ref{G-Breaking}). For example we have
\begin{eqnarray}\label{E-components1}
&& F_{\tau z}=A_{\tau,z}-A_{z,\tau}\cr
&& \;\;\;\;\;\;=\partial_z(\sqrt{g_{00}}
\partial_z\psi)-\partial_\tau(-\sqrt{g_{00}}g^{\tau\tau}
\partial_\tau\psi)\cr
&& \;\;\;\;\;\;=\partial_z(\sqrt{g_{00}}
\partial_z\psi)-\omega^2 \sqrt{g_{00}}g^{00} \psi,
\end{eqnarray}
where we have used the facts that  $A_\tau=\epsilon_{\tau
z}\partial^z \psi$  and $A_z=\epsilon_{z\tau}\partial^\tau\psi$.
Similar  manipulations lead to  other components of Maxwell tensor as follows
\begin{eqnarray}\label{E-components2}
&& F_{\tau x}=ik_x \sqrt{g_{00}}\partial_z\psi-k_y\omega\phi \cr
&& F_{\tau y}=ik_y \sqrt{g_{00}}\partial_z\psi+k_x\omega\phi \cr
&&F_{xy}=(\epsilon_{xy}\partial^y\phi)_{,y}-(\epsilon_{yx}\partial^x\phi)_{,x} = -k_x^2\phi-k_y^2\phi=-k_\bot^2\phi \cr
&& F_{zx}=-ik_y\partial_z\phi-k_x\omega g^{00}\sqrt{g_{00}}\psi\cr
&& F_{zy}=-ik_x\partial_z\phi-k_y\omega g^{00}\sqrt{g_{00}}\psi.
\end{eqnarray}
Using the above results we have
\begin{eqnarray}\label{E^2-1}
&&|\mathbf{E}|^2=|\partial_z(\sqrt{g_{00}}
\partial_z\psi)-\omega^2 \sqrt{g_{00}}g^{00} \psi|^2 + \cr
&& \;\;\;\;\;\;\;\;\;\;\;\;\;\;\;\;\;\;\;\;\;\;\;\;\;\;\;\;\;|ik_x \sqrt{g_{00}}\partial_z\psi-k_y\omega\phi|^2 +
|ik_y \sqrt{g_{00}}\partial_z\psi+k_x\omega\phi|^2.
\end{eqnarray}
Taking the vacuum expectation value of the above quantity, namely $<0||\mathbf{E}|^2|0>$, it should be noted that terms like
$\phi\partial_z\psi$ have no contribution  as the creation and annihilation operators appearing
in such terms belong to two different spaces and thereby leading to zero contribution. Ignoring such terms
in the expansion of $|\mathbf{E}|^2$ we end up with
\begin{eqnarray}\label{E^2-2}
&&|\mathbf{E}|^2=|\partial_z(\sqrt{g_{00}}
\partial_z\psi)-\omega^2 \sqrt{g_{00}}g^{00} \psi|^2 + g_{00}
k_\bot^2|\partial_z\psi|^2 + k_\bot^2\omega^2|\phi|^2.
\end{eqnarray}
Using the fact that $\psi$ satisfies the Klein-Gordon equation, $\nabla^\mu\nabla_\mu \psi=0$, we have
\begin{eqnarray}\label{KG for psi}
\partial_z(\sqrt{g_{00}}\partial_z\psi)-\omega^2 \sqrt{g_{00}}g^{00} \psi=k_\bot^2
\sqrt{g_{00}}\psi,
\end{eqnarray}
which on substitution for the first term in the right hand side of equation (\ref{E^2-2}) leads to
\begin{eqnarray}\label{E^2-final}
&&|\mathbf{E}|^2 = k_\bot^4g_{00}|\psi|^2 + g_{00}
k_\bot^2|\partial_z\psi|^2+k_\bot^2\omega^2|\phi|^2.
\end{eqnarray}
In the same way for the square of the magnetic induction we have
\begin{eqnarray}\label{B^2-1}
&&|\mathbf{B}|^2=k_\bot^4|\phi|^2+|-ik_y\partial_z\phi-k_x\omega
g^{00}\sqrt{g_{00}}\psi|^2+|-ik_x\partial_z\phi-k_y\omega
g^{00}\sqrt{g_{00}}\psi|^2
\end{eqnarray}
manipulating in the same way as for the square of the electric field we end up with
\begin{eqnarray}\label{B^2-final}
&&|\mathbf{B}|^2=k_\bot^4|\phi|^2+k_\bot^2|\partial_z\phi|^2 + k_\bot^2\omega^2g^{00}|\psi|^2
\end{eqnarray}
Now substituting the above results for the squares of the electric and magnetic fields in $T_{00}$ we have
\begin{eqnarray}\label{Energy-Momentum-1}
|\mathbf{E}|^2 + g_{00}|\mathbf{B}|^2 = k_\bot^4g_{00}|\psi|^2 + g_{00}
k_\bot^2|\partial_z\psi|^2+k_\bot^2\omega^2|\phi|^2\cr
+ g_{00}k_\bot^4|\phi|^2 + g_{00}k_\bot^2|\partial_z\phi|^2v + g_{00}k_\bot^2\omega^2g^{00}|\psi|^2\cr
= k_\bot^2[(\omega^2 + g_{00}k_\bot^2)|\phi|^2 + g_{00}|\partial_z\phi|^2] + k_\bot^2[(\omega^2 + g_{00}k_\bot^2)|\psi|^2 + g_{00}|\partial_z\psi|^2]
\end{eqnarray}
It is easy to see that the expectation value of the two terms in the brackets are nothing but the expectation values of the time-time components of  energy-momentum tensors of the scalar fields $\phi$ and $\psi$ respectively. To show this we note that
\begin{eqnarray}\label{Energy-Momentum-Scalar}
&&T_{00}^{scalar}=\phi_{,0}\phi_{,0}^*-\frac{1}{2}g_{00}[g^{00}\phi_{,0}\phi_{,0}^*+\delta^{ij}\phi_{,i}\phi_{,j}^*]\cr
&&=\frac{1}{2}\phi_{,0}\phi_{,0}^* + \frac{1}{2}g_{00}[|\partial_z\phi|^2+(ik_x)(-ik_x)|\phi|^2+(ik_y)(-ik_y)|\phi|^2]\cr
&&=\frac{1}{2}(i\omega)(-i\omega)|\phi|^2 + \frac{1}{2}g_{00}|\partial_z\phi|^2 + \frac{1}{2}g_{00}k_\bot^2|\phi|^2,
\end{eqnarray}
so that for the scalar fields in our problem
\begin{eqnarray}\label{Energy-Momentum-ScalarFinal}
T_{00}^{\phi}=\frac{1}{2}(\omega^2 + g_{00}k_\bot^2)|\phi|^2 + \frac{1}{2} g_{00}|\partial_z\phi|^2\\
T_{00}^{\psi}=\frac{1}{2}(\omega^2 + g_{00}k_\bot^2)|\psi|^2 + \frac{1}{2} g_{00}|\partial_z\psi|^2.
\end{eqnarray}
substituting these relations in (\ref{Energy-Momentum-1}), the vacuum expectation value of the time-time component of the 
energy-momentum tensor is given by
\begin{eqnarray}\label{Energy-Momentum-total}
<0|T_{00}^{total}|0>=\frac{1}{2}<0|(|\mathbf{E}|^2 + g_{00}|\mathbf{B}|^2)|0> = k_\bot^2(<0|T_{00}^{\phi}|0> + <0|T_{00}^{\psi}|0>).
\end{eqnarray}
Note that one could absorb the factor $k_\bot^2$ into the definitions of $\phi$ and $\psi$. Now from our previous study on the energy
density of scalar fields in Fermi metric \cite{NouriNazari}, based on the spacetime index of refraction, we know that
\begin{eqnarray}\label{Energy-Momentum-total-1}
&&<0|T_{00}^{\phi}|0> = <0|T_{00_{flat}}^{\phi}|0> (1+\frac{\textrm{g}\ell}{2})\\
&&<0|T_{00}^{\psi}|0> = <0|T_{00_{flat}}^{\psi}|0>(1+\frac{\textrm{g}\ell}{2}).
\end{eqnarray}
Substituting these relation back into the equation (\ref{Energy-Momentum-total}) we have
\begin{eqnarray}\label{Energy-Momentum-total-final}
<0|T_{00_{curved}}^{EM}|0>=(<0|T_{00_{flat}}^{\phi}|0> + <0|T_{00_{flat}}^{\psi}|0>)(1+\frac{\textrm{g}\ell}{2})\cr =  <0|T_{00_{flat}}^{EM}|0>(1+\frac{\textrm{g}\ell}{2}).\;\;\;\;\;\;\;\;\;\;\;\;\;\;\;\;\;\;\;\;\;\;\;\;\;\;\;\;
\end{eqnarray}
In effect this last relation shows that one could have obtained the vacuum electromagnetic energy density of a Casimir apparatus in a Fermi metric by assigning an index of refraction to the underlying spacetime and thereby applying the conjecture introduced in the introduction on the frequency of the
virtual photons as the content of the quantum vacuum.
\section{discussion}
Using the fact that one could employ the gauge transformation and the Lorentz condition on a vector potential to decompose it formally into two scalar fields satisfying the Klein-Gordon equation, it was shown how the boundary conditions on the vector potential are transformed to those on the two scalar fields. Considering the same problem in the context of a Casimir apparatus in a weak gravitational field whose vacuum content are virtual photons propagating between the two plates, it was shown how the above decomposition leads to the same frequency shift on virtual photons as was found previously for virtual scalar particles as the constituents of vacuum. Since we have formally two scalar fields replacing an electromagnetic field, it may be expected that the energy content to be doubled compared to the case of Casimir vacuum composed of virtual massless scalar particles. This expectation is not a trivial one since the decomposition of the electromagnetic field into two scalar fields  in Fermi metric is such that they are defined in the two totally different
sectors of the space time. Despite this fact, it was shown explicitly that the two scalar fields have equal contributions to the energy density leading to the expected energy content for the electromagnetic field as the constituent of the vacuum in a Casimir apparatus in a weak gravitational field. This calculation was carried out without any resort to the Green's functions method frequently employed in the literature.
\section *{Acknowledgments}
The authors would like to thank University of Tehran for supporting
this project under the grants provided by the research council. They
also thank Center of excellence on the structure of matter and particle interactions of the
University of Tehran for partial support.
\pagebreak
\appendix
\section{Note on the scalar functions $\phi$ and $\psi$}
Here we first show that $\psi$ and $\phi$ satisfy the Klein-Gordon
equation and then derive relations (\ref{A-mode1})-(\ref{A-mode4}).
For example from equations (\ref{A-form}) and (\ref{G-Breaking}) we have
\begin{eqnarray}\label{apx1}
A_a=\epsilon_{ab}\nabla^b\psi=f_a^{(\lambda)}(z)\exp^{-i\omega
t+i\textbf{k}_\bot.\textbf{x}_\bot},
\end{eqnarray}
which is consistent with the choice of $\psi$ (and also  $\phi$ ) according to
the general form (\ref{SF0}). From the above equation and equation (\ref{ME}) (in the absence of curvature term), 
for the component  $A_3$ of the vector field we have, 
\begin{eqnarray}\label{apx2}
\square A_3 = \square (\epsilon_{30}\nabla^0\psi) = -i\omega \square
(\epsilon_{30} g ^{00}\psi)=0,
\end{eqnarray}
in which we have used the general form (\ref{SF0}) for $\psi$ in the last equation. Since the
covariant derivative of metric vanishes we have
\begin{eqnarray}\label{apx3}
\square (\epsilon_{30} g^{00}\psi) = \square (-\frac{1}{\sqrt{g_{00}}}\psi) = 0 \Rightarrow \square \psi=0
\end{eqnarray}
in other words $\psi$ satisfies the Klein-Gordon equation.
In terms of the scalar functions $\phi$ and $\psi$ one can choose the 
following simplest linearly independent mode functions for $A^{(\lambda)}_\mu$ \cite{Higuchi}
\begin{eqnarray}\label{apx4}
A_\mu ^{(0)}=(\partial_0,\partial_3,0,0)\phi\\
A_\mu^{(1)}=(p_0,p_3,0,0)\psi\\
A_\mu^{(2)}=(0,0,p_1,p_2)\phi\\
A_\mu^{(3)}=(0,0,\partial_1,\partial_2)\phi,
\end{eqnarray}
in which the first and last modes are pure gauge modes and the physical ones $A_\mu^{(1)}$ and $ A_\mu^{(2)}$ 
reduce to those in the text. For example, the non-zero components of $A_\mu^{(1)}$ are given by 
\begin{eqnarray}\label{apx5}
p_0\psi=\epsilon_{01}\nabla^1\psi=\epsilon_{01}g^{11}\partial_1\psi=(1+gz)\partial_z\psi\cr
p_3\psi=\epsilon_{30}\nabla^0\psi=\epsilon_{30}g^{00}\partial_0\psi=-i(1 - gz)\omega\psi,
\end{eqnarray}
leading to the mode function (\ref{A-mode2}). 
\pagebreak

\end{document}